 \documentstyle[12pt]{article}

   \let\d=\delta \let\e=\epsilon
    \let\k=\kappa
\let\l=\lambda \let\m=\mu \let\n=\nu   \let\r=\rho
\let\s=\sigma   \let\f=\phi  
\let\w=\omega       \let\D=\Delta  \let\L=\Lambda
 \let\P=\Pi   \let\F=\Phi 
 
 \let\ci=\cite 
  
 \def\bd{\begin{document}} \def\ed{\end{document}}
\def\ds{\documentstyle} \let\fr=\frac \let\bl=\bigl \let\br=\bigr
\let\Br=\Bigr \let\Bl=\Bigl
\let\bm=\bibitem
\let\na=\nabla
\let\pa=\partial \let\ov=\overline
\newcommand{\be}{\begin{equation}}
\newcommand{\ee}{\end{equation}}
\def\ba{\begin{array}}
\def\ea{\end{array}}
\newcommand{\bea}{\begin{eqnarray}}
\newcommand{\eea}{\end{eqnarray}}
\newcommand{\ra}{\rightarrow}
\newcommand{\lra}{\longrightarrow}
\newcommand{\Lra}{\Leftrightarrow}

\newcommand{\tamphys}{Center for Theoretical Physics\\
Physics Department \\ Texas A \& M University
\\ College Station, Texas 77843, USA}

\begin{document}

\hfill{CTP-TAMU-50/92}

\vspace{24pt}

\begin{center}
{ \large {\bf BRS Operators and
Covariant Derivatives in Loop Space
for p-Branes coupled to Yang-Mills}}

\vspace{36pt}

J. A. Dixon

\vspace{6pt}

{\tamphys}

\vspace{24pt}

June, 1992

\vspace{48pt}

\underline{ABSTRACT}

\end{center}
Canonical forms are given for the nilpotent BRS operator $\d$ and the
covariant `loop space' derivative ${\cal D}_{\m}$ for the p-brane
fields for all odd p. The defining characteristic of ${\cal D}_{\m}$ is
that it is a functional derivative operator which generalizes the ordinary
functional derivative and also commutes with $\d$. Methods of construction
for the canonical forms are discussed.

\vfill

\baselineskip=24pt

\pagebreak

\setcounter{page}{1}

\section{ Introduction}

The existence of two alternative formulations  of 10-dimensional
supergravity-Yang-Mills field theory, suggests that there may exist\ci{duff}
a fundamental heterotic super-5-brane theory that is dual
\ci{strominger}\ci{dufflu}
to the heterotic string (1-brane).  These two
p-brane (p=5 and p=1) theories should couple to the two dual versions
 (with
antisymmetric tensors $B_6$ and $B_2$ respectively) of the
supergravity-Yang-Mills theories. The action for the heterotic
superstring
coupled to this supergravity-Yang-Mills background is known (See
\ci{bergs} for
the formulation closest to our present needs). The bosonic part of the
corresponding action for the 5-brane coupled to Yang-Mills
has been discussed in \ci{dds},
 but this
has not yet been made supersymmetric. On the other hand,
the actions for
supersymmetric p-branes that are not coupled to Yang-Mills
are known\ci{bst}.

One way to get the equations of motion of the $B_2$ theory is to derive
it from
light-like integrability applied to the covariant derivatives ${\cal
D}_{\m}(\s)$ in loop superspace\ci{bergs}.  Hence one might be able to
 make some
progress in finding the heterotic supersymmetric 5-brane by examining
the
analog problem for the 5-brane.  However, the present discussion is
restricted to bosonic fields  only.

In this paper, the formulation of the BRS operator and
 the
covariant derivative for the coupling of Yang-Mills to the p-branes for
 odd p are discussed in a general way.
The operators for even p will no doubt involve products of  the odd p
structures as happens in \ci{dds}.  In general, this formulation involves
complicated polynomials in the Yang-Mills field $A_{\m}^a$ and a
composite
field $K_{i}^a$. In this paper no new examples of these polynomials are
 given,
but the way they arise is explained and it is proved that they always
exist.  Various methods for their construction are discussed.
A summary of the results is given  in the
conclusion.  The notation here follows that of \ci{dd}.

\section{Covariant Functional Derivatives}

How should one define a covariant derivative?  What properties should
it have?  One way to find a covariant
derivative is to express the
velocity in terms of the Hamiltonian variables, and then
  quantize the theory.  For example in  electromagnetism,
one gets the covariant derivative:
\be
{\cal D}_{\m} = {\dot X}_{\m}
= p_{\m} + A_{\m}
=i \left [ {\pa \over \pa X^{\m}} - i A_{\m} \right ]
\label{em}
\ee
Another way  is to define the covariant
derivative to be a generalization of the ordinary derivative
which commutes with the BRS operator $\d$. These two ways are
probably equivalent, and both are used for the string
in \ci{bergs}.

A covariant derivative should have several
properties, and these can best be enumerated
by considering
a simple example in field theory, which
also turns out to be relevant to our problem below.  Suppose that $\f^i$ is
a scalar field coupled to  Yang-Mills theory.  The covariant
derivative can be written in the following way:
\be
D_{\m j}^{i } \f^j =
 \pa_{\m} \f^i + A^a_{\m} T^{a i}_{\; \; j } \f^j
 \ee
where the matrices $T^a$ satisfy:
\be
[ T^a, T^b ] = f^{abc} T^c
\ee
The covariant derivative can be obtained by defining a
functional derivative operator as follows:
\be
{\cal D}_{\m }
=
\int \; d^Dx \; \left \{ D_{\m j}^{i } \f^j (x) \fr{\d}{\d \f^i(x)}
\right \}
\ee
Using this expression for ${\cal D}_{\m }$ one finds that:
\be
[{\cal D}_{\m} \f ]^i = D_{\m j}^{i } \f^j
\ee
Now note the following features of this functional
derivative form of the covariant derivative:
\begin{enumerate}
\item
\label{der}
${\cal D}_{\m}$  is a functional derivative operator which is
integrated over spacetime and summed over all indices except
the one uncontracted spacetime index ${\m}$.  The
advantage in writing
${\cal D}_{\m}$  in this way is that it makes  ${\cal D}_{\m}$
into a derivation on the algebra of polynomial
functionals of the   field $\f^i$, so that the covariant
derivative behaves much like an ordinary derivative.
When acting on invariant local polynomials in $\f^i$, like
say $d_{ijk} \f^i \f^j \f^k $, where $d_{ijk}$ is an invariant
tensor, ${\cal D}_{\m}$  yields the same result as  the ordinary
derivative.
\item
In this functional derivative form, one can verify  that ${\cal D}_{\m}$
commutes with the BRS operator: \be
[{\cal D}_{\m}
, \d  ] =0
\label{commutes}
\ee
where the anticommuting BRS operator $\d$ is:
\be
  \d
= \int \; d^Dx \; \left \{ D_{\m}^{ab} \w^b \fr{\d}{\d A_{\m}^a(x)}
- \fr{1}{2} f^{abc} \w^b \w^c \fr{\d}{\d \w^a(x)}
-
 T^{a i}_{\; \; j } \f^j \w^a  \fr{\d}{\d \f^i(x)} \right \}
\label{brsf}
\ee
Note that
\begin{enumerate}
\item
$\d$ is an anti-derivation (an anticommuting derivation),
\item
$\d$ is a functional derivative
operator integrated over space-time and summed over all indices,
\item
$\d$ is nilpotent:
\be
\d^2 =0.
\ee
\end{enumerate}
\item
Because of the above properties, the  (multiple) commutators:
\be {\cal D}_{ \m_2 \m_1 } =
 [{\cal D}_{\m_2}
,  {\cal D}_{\m_1 } ]
\ee
\be {\cal D}_{ \m_k \cdots \m_1 } =
 [{\cal D}_{\m_k}
,  {\cal D}_{\m_{k-1} \cdots \m_1 } ]  ; \; \; (k \geq 3)
\ee
also satisfy
\be [ {\cal D}_{ \m_k \cdots \m_1 }
, \d  ] = 0
\ee
For example, in this case,  one gets:
\be
  {\cal D}_{ \m \n  }
= \int \; d^Dx \; \left \{  -
 {F}^a_{\m \n} T^{a i}_{\; \; j } \f^j
\fr{\d}{\d \f^i(x)} \right \}
\ee
Here the Yang-Mills curvature is
\be
F_{\m \n}^a = \pa_{\m} A_{\n}^a - \pa_{\m} A_{\n}^a
+ f^{abc} A_{\m}^b A_{\n}^c
\ee
 \item
In this functional form, one can add covariant derivatives for different
fields, and this will generally change the value of these commutators
above. This is particularly true if one adds `covariant derivatives'
for gauge fields.  For example, one could take:
\be
{\cal D}_{\m \; \rm new}
=
\int \; d^Dx \; \left \{ D_{\m j}^{i } \f^j (x) \fr{\d}{\d \f^i(x)}
+ {1 \over 2}  {F}^a_{\m \n} \fr{\d}{\d A^a_{\n}} \right \}
\label{newcov}
\ee
which still satisfies (\ref{commutes}),
but then quite a different result for ${\cal D}_{ \m \n  }$,
is obtained, namely:
\be   {\cal D}_{ \m \n \; \rm new}  =
- {3 \over 4}
\int \; d^Dx \;
D_{\r}^{ab} F^b_{\m \n} {\d \over \d A^a_{\r} }
\ee
Only the gauge part of the commutator remains--the ${\d \over \d \f} $
part is gone.

\end{enumerate}

\section{ Covariant Derivative for String}

For the reasons above, we will look for a Yang-Mills gauge covariant
derivative  for a p-brane that is a
functional derivative operator ${\cal D}_{\m}$ that commutes with
the nilpotent BRS functional derivative operator $\d$:
\be
[ \d, {\cal D}_{\m} ] = 0
\ee
In \ci{dd}, the functional derivative
form of the BRS operators $\d$ for the
string and the 3-brane was discussed, and those results will be used here.
The results for the string were based on the analysis in \ci{bergs}.
For the string, the result in \ci{dd} was that the BRS operator is:
\be
\d = \d_{\rm fields } + \d_{{\rm string}}
\ee
and
\[
\d_{\rm fields }
= \int \; d^Dx \; \left \{ D_{\m}^{ab} \w^b \fr{\d}{\d A_{\m}^a(x)}
- \fr{1}{2} f^{abc} \w^b \w^c \fr{\d}{\d \w^a(x)} \right.
\]
\[
+
[ \pa_{[\m} \L_{\n]}
+ n \pa_{[\m} \w^a A_{\n]}^a]  \fr{\d}{\d B_{\m \n}(x)}
+
[ - \pa_{\m} B + n \w^a \pa_{\m} \w^a  ] \fr{\d}{\d \L_{\m}(x)}
\]
\be \left.
+{1\over6}  n f^{abc} \w^a \w^b \w^c \fr{\d}{\d B } \right \}
\label{func1}
\ee
\be
\d_{\rm string} =
\prod_{\s' \n, m} \int dX^{\n}(\s') dy^m(\s')
\left \{
\left (
\int d \s \left [
- \w^a T_a(\s)
+ \L_{\m} \fr{ d X^{\m} }{ d \s} \right ] \F
\right )
 {\d \over \d \F} \right \}
\label{func2}
\ee

Here $\F$ is the string field, which is to be considered an
arbitrary functional of the functions $X^{\m}(\s)$ and
$y^{m}(\s)$ in the same sense that in field theory
the scalar fields $\f^i$ are
arbitrary  functions of
$x$.  $X^{\m}(\s)$ and  $y^{m}(\s)$
are to be considered a set of D  and ${\rm Dim(G)}$ arbitrary
functions of the string parameter $\s$, respectively.
The functional $\F$ has no explicit $\s$ dependence since it is
integrated over many copies of the $\s$ variables--one for each
variable $X$ or $y$ in it. One could write it in the form
(ignoring the $y$ variables):
\be
\F[X] = \sum_n
\prod_{i=1}^n \int d \s_i \F_{\m_1 \cdots \m_n}(\s_1 \cdots \s_n)
X^{\m_1}(\s_1)
\cdots
X^{\m_n}(\s_n).
\ee
Thus the integrated $\L$ term above is simply
a factor which multiplies
$\F$, whereas the $T^a$ term above is a functional derivative operator
which operates on the $y$ variables in
$\F$.

For the covariant derivative on the string functional $\F$, we
want an operator that commutes with the above $\d$.  A
form which satisfies this equation for the string   is
given by:
\[
{\cal D}_{ \; {\rm string}} =
\prod_{\s' \r, m} \int dX^{\r}(\s') dy^m(\s')
\left [
\int d \s V^{\m}[X(\s)] {\cal D}_{\m }(\s) \F
\right ] {\d \over \d \F}
\]
\[
=
\prod_{\s' \r, m} \int dX^{\r}(\s') dy^m(\s')
\]
\be
\left (
\int d \s V^{\n}[X(\s)]
\left [
\fr{\d}{\d X^{\n}(\s)} + A_{\n}^a T_a(\s)
+ [  n   A^a_{\n} A^a_{ \l} - 2  B_{\n \l} ]
\fr{ d X^{\l} }{ d \s} \right ]
\F
\right )
 {\d \over \d \F}
\label{string}
\ee
 Here $V^{\n}[X(\s)]$ can be an arbitrary function of
$X(\s)$ and it could have more indices if desired.  It cannot be a
function of $y(\s)$ however. This operator
${\cal D}_{\m \;  { \rm string}}(\s)$
was given (in a different form) in \ci{bergs}, where it was deduced both
from the requirement that
$[ \d, {\cal D}_{\m \;  { \rm string}}(\s) ] = 0$ and
  from the heterotic Green-Schwarz
superstring action using
canonical quantization as in (\ref{em}) above.
Using the Kac-Moody
 algebra
obeyed by the $T^a$ operators:
\be
[ T^a(\s), T^b(\s') ] = f^{abc} T^c(\s) \d(\s - \s')
+ 2 n \d^{ab} { d \over d \s} \d(\s - \s'),
\ee
it is straightforward to show that:
\be
[\d,  {\cal D}_{\m}(\s) ] =
[ \d_{{\rm string}}, {\cal D}_{\m \; {\rm string}}(\s)  ]
+
[ \d_{\rm fields },{\cal D}_{\m \; {\rm string}}(\s)  ] =0
\ee

\section{BRS operators for p-branes}

In \ci{dd}, the BRS operator for the 3-brane was discussed in
detail.   Here form notation will be used, since it is more convenient.
The operator $\d$ should however be thought of as a functional
derivative operator as in (\ref{func1})
and (\ref{func2}).
The generalization of the results in
\ci{dd}  to the case of arbitrary p is:
\be \d   \F  =
   \int  \left [ -
   \w^a T^a(\s)  d^p \s
+ n C^1_p
+ \L_p^1  \right ] \F
\label{genBRS}
\ee
\be
\d   A^a =
-d \w^a - f^{abc} A^b \w^c
 \ee
\be
\d \w^a = - {1\over2} f^{abc} \w^b \w^c
\ee
\be
\d B^0_{p+1}
 = n I^1_{p+1} +  d \L^1_{p }
\label{deltaB}
\ee
\be
\d \L^1_{p} = n I^2_p  +
d B^2_{p-1}
\ee
\[
\cdots
\]
\be
\d B^{p+1}_0 = n I^{p+2}_0
\label{last}
\ee
Here   $T^a$ generates
 Yang-Mills transformations on $K_i^a[y(\s)]$:
\be \d K^a(\s')
=
\left [  \int \w^b(\s) T^b(\s) d^p \s , K^a(\s') \right ]
= - d \w^a(\s') - f^{abc} K^b(\s') \w^c(\s')
\label{transk}
\ee
Here it is assumed that the expressions depend only on the
parameters indicated below:
\be
C^1_p(\s)
=
C^1_p\Bigl ( A[X(\s)], \w[X(\s)], K[y(\s)]   \Bigr )
\ee
\be
I^i_{p +2 -i}[A,\w](\s)
=
I^i_{p +2 -i}\Bigl ( A[X(\s)], \w[X(\s)] \Bigr )
\ee
\be
I^i_{p +2 -i}[K,\w](\s)
=
I^i_{p +2 -i}\Bigl ( K[y(\s)], \w[X(\s)] \Bigr )
\ee
\be
T^a(\s) =
T^a[y(\s), {\d \over \d y(\s) } ]
\ee
Note that on the right hand sides of equations
(\ref{genBRS}-\ref{last}),
only  $T^a(\s)$ is an operator--the rest of the
quantities are just numbers, once field values are assigned.
The expressions $I^i_{p +2 -i}$ appear in the descent equations
of Yang-Mills theory:
\be
\d I^i_{p +2 -i} =
d  I^{i+1}_{p +1 -i} \; \; i = 0, 2 \cdots {p+1}.
\label{descent}
\ee
\be
\d I^{p+2}_{0} = 0
\ee
Following the analysis in \ci{dd}, we will assume that
the $y$-dependent operators $T^a$ satisfy  the following
algebra:
\[
 \int d^p \s \w^a(\s) T^a(\s)  \int d^p \s'  \w^b(\s') T^b(\s')    =
\]
\be
\int \left [
\fr{1}{2}  f^{abc} \w^a(\s) \w^b(\s) T^c(\s)   d^p \s
+ n  I^2_p(K,\w) \right ]
\ee
Note that here $I^2_p$ is a function of
 $K$ and not   $A$, in accord with the fact that $T^a$ depends
only on the $y$ variables, as does $K$.  In a Hamiltonian
treatment, these variables $K$ should arise from the action
given in \ci{dds}, and can identified with the spatial
parts of the $K^a_i$ defined there.
The above operator (\ref{genBRS}-\ref{last})
satisfies  $\d^2 =0$ if:
\be
\d C^1_p(A, K,\w)=   I^2_p(A,\w)  -   I^2_p(K,\w)
 + d C^2_{p -1}(A, K,\w)
\label{omega}
\ee
where the argument $\s$ is understood everywhere.
Here $C^2_{p -1}(A, K,\w)$ is a new function which arises because
nilpotence requires only that the integral of this expression be zero.

But equation (\ref{omega}) is just a descent from the equation that was
solved in \ci{dds}.  The    general form
of these descent equations is:
\[
\d C^{i-1}_{p+2-i}(A, K,\w)
=  I^i_{p+2-i}(K,\w)
\]
\be
-   I^i_{p+2-i}(A,\w)
+  d C^{i}_{p+1-i}(A, K,\w);\;\; (i \geq 1)
\label{bigdes}
\ee
In the above, the operator $\d$ performs Yang-Mills transformations on
$A$ and $K$ and can be written in the form of
equations (\ref{afirst}--\ref{alast}) below.
In \ci{dds}, a formula was given for
$C^{0}_{p+1 }(A, K )$, which satisfies this equation for
$i=1$.  In that paper, we ignored the term
$C^{1}_{p}(A, K,\w)$ since we were only interested in the
action there.  Evidently one can obtain $C^{1}_{p}$
from the results in \ci{dds}.  Note that in (\ref{bigdes}), the
expression $d C^{i}_{p+1-i}(A, K,\w) $ generally will not be zero
because  acting on (\ref{bigdes}) with $\d$ gives:
\be
0
= d I^{i+1}_{p+1-i}(K,\w)
-  d I^{i+1}_{p+1-i}(A,\w)
- d \d C^{i}_{p+1-i}(A, K,\w);\;\; (i \geq 1)
\label{bigdes1}
\ee
and since the first two terms are not zero, and do not cancel
in general, the third term is not zero.  When the descent arrives at
terms which are not dependent on the gauge fields, but only on the
ghost $\w$, the terms will become zero.

Better   methods,
probably using the `Russian Formula'\ci{msz},
to find $ C^{1}_{p}(A, K,\w) $ are doubtless
available, but this paper is  concerned only with
the existence of $C^{1}_{p}$, which follows from the above.

The two simplest examples of $C^{1}_{p}$ are, for the string,
\be
C^1_1 = 0
\ee
and for the 3-brane (see \ci{dd}),
\be
C^1_3 = d^{abc} K^a A^b d \w^c
\ee
It is easy to verify equation (\ref{omega}) for these two cases,
using
\be
I^2_3(A,\w) = d^{abc} A^a d \w^b
d \w^c
\ee
To go beyond the 3-brane requires the
general form of $I^2_p(K,\w)$ for p
odd. It will be shown below that this can be written in the form:
\be \int_{\s}
I^2_{p}(K,\w)
= \int_{\s}
 \w^{a} d \w^{b}
M_{p-1}^{(a  b )}[K]
\label{MMM}
\ee
For $p=1$ we have
\be
M_0^{ab} = \d^{ab}.
\ee
and for $p \geq 3$:
\be
M_{p-1}^{a b} = d N_{p-2}^{ab}
\ee
So it is true for all odd $p$ that:
\be
d M_{p-1}^{ab} = 0
\ee
This can be seen from the following expressions, which,
for $p \geq 3$, were derived in \ci{zumino}:
\be
I^2_1 = \w^a d \w^a
\ee
\[
I^i_{p+2-i}(K,\w)
=
{ (p+3)(p+1)\cdots (p+3-2i) \over
2^{i+1} i!}
\]
\[
\int_0^1 dt (1-t)^i  P[ (d \w)^i, K, (t dK + t^2 K^2 )^{p+1-2i \over 2} ]
\]
\be =
d \w^{a_1} \cdots d \w^{a_i}
N^{a_1 \cdots a_i}[K]; \;
( p \; {\rm odd});\;  (p \geq 3) ;\;   (i \leq {p+1 \over 2})
\label{MM}
\ee
where $N^{a_1 \cdots a_i}[K]$
is defined by the above and the notation $P$ is defined by:
\be
P[ H_1 , H_2 , (H_3)^n]
\equiv
d^{(a b a_1 \cdots a_n)}
H_1^a H_2^b H_3^{a_1} \cdots  H_3^{a_n}
\ee
for any Lie-algebra valued forms $H_j$. Here
\be
(K^2)^a \equiv {1\over 2} f^{abc} K^b K^c
\ee
Removing the $\w$ fields by functional differentiation
and replacing $N$ by its dual
yields the form:
\be
\left [ T^a(\s), T^b(\s') \right ]
= f^{abc} T^c(\s) \d^p(\s-\s')
+ n \pa_i
{\tilde N}^{(ab)[ij]}(\s) { \pa \over \pa \s'^j} \d^p(\s-\s')
\ee
where
\be
N^{(ab)}_{p-2}
=
N^{(ab)}_{i_1 \ldots i_{p-2} } d \s^{i_1}
\cdots d \s^{i_{p-2}}
\ee
\be
{\tilde N}^{(ab)[i_1 i_2]}(\s)
=
\e^{i_1 \ldots i_p}
N^{(ab)}_{i_3 \ldots i_{p} }(\s)
\ee
for $p \geq 3$.
This is the general form of the Mickelsson-Faddeev algebra
for arbitrary odd p.

\subsection{  Covariant Derivatives for p-branes}
Now we want to generalize the covariant derivative from
the string case to the case of the p-brane for
general odd p.  We continue to think of the operator in the
form (\ref{string}), but shall use a simpler notation by
applying the operator to the p-brane functional $\F$ and
removing the integration over $\s$.

The following is the canonical form:
\be
 {\cal D}_{\m}(\s) \F  =
  \left \{
\fr{\d}{\d X^{\m}(\s)} + A_{\m}^a(\s) T_a(\s)
 +      n  J^0_{\m p }(\s) - (p+1)  B_{\m p}(\s)
   \right \}   \F
\label{gencov}
\ee
where the new function $J$ is a function of the fields:
\be
J^0_{\m p }(\s)=
J^0_{\m p }\left [ A[X(\s)],  K[y(\s)] \right ]
\ee
and has the form:
\be
J^0_{\m p } =
J^0_{\m \m_1 \cdots \m_p } \P^{\m_1 \cdots \m_p}
\ee
as does $B_{\m p}$:
\be
B_{\m p } =
B_{\m \m_1 \cdots \m_p } \P^{\m_1 \cdots \m_p}
\ee
Covariance of this  covariant derivative requires
that it satisfy
\be
[\d, {\cal D}_{\m}(\s)]=0
\label{comm}
\ee
which
implies that:
\[ \left [ \d ,  J^0_{\m p }[A,K ](\s) \right ]  =
-  (p+1)
  I^1_{\m p}[A, \w](\s)
\]
\[
-  \left [ \left ( {\d \over \d X^{\m}(\s) }
+ A_{\m}^a(\s) T_a(\s) \right ),
\int_{\s'}  C^1_{ p }(\s') \right ]
\]
\be
+ A_{\m}^a(\s) d \w^b(\s) M_{p-1}^{(ab)}(\s)
\equiv  J^1_{\m p }[A,K,\w](\s)
\label{coveq}
\ee
where we have used equations (\ref{genBRS},\ref{MMM},\ref{gencov})
and have defined a new quantity  $J^1_{\m p }[A,K,\w](\s)$.
This is an equation which determines $J^0_{\m p}[A,K ]$
in terms of $I^1_p[A,\w]$, $C^1_p[A,K,\w]$ and $M_{p-1}^{(ab)}[ K ]$.
Why should a $J^0_{\m p}[A,K ]$ exist
which satisfies this equation?    Note that (\ref{comm})
 implies the
relation: \be
\left [\d , J^1_{\m p}[A,K,\w] \right ]=0
\label{comb}
\ee
It is easy to verify that
the $\L$ dependence disappears from $J^1_{\m p}$, so that
indeed $J^1_{\m p}$ depends only on the variables
$A$, $K$ and $\w$.  Then it is also evident that in equation
(\ref{comb}), only the transformations of
$A$, $K$ and $\w$ are relevant.
These transformations are summarized and used in the equations
(\ref{afirst}-\ref{alast}) in the Appendix below.
However note that $J^1_{\m p}[A,K,\w]$ actually arises
from $\d$ transformations including (\ref{deltaB})
and (\ref{genBRS}), so that
it is not at all obvious that a $J^0_{\m p}[A,K ]$ which
is only a function of $A$, $K$ and $\w$ and which
satisfies (\ref{coveq})
should exist--all we know so far is that some function of
a larger set of variables with a larger transformation rule
exists which gives rise to $J^1_{\m p}[A,K,\w]$.

In appendix A it is shown that
the local unintegrated cohomology of the BRS operator
restricted to the fields $A,K,\w$ is trivial in the ghost
charge one sector.  It then follows that $J^0_{\m p}$ exists with
the desired properties.

It is instructive to see how this works  for the
string:
\be
M_1^{(ab)} = \d^{ab}
\ee
 We already know from (\ref{string}) the form of $J^0_{\m 1}$ for the string:
\be
J^0_{\m 1} =  A^a_{\m} A^a_{\n} {\pa X^{\n} \over \pa \s}
\ee
and it is easy to verify that (\ref{coveq}) holds for
this case.  The form of $J^0_{\m 3}$ for the 3-brane is
much more complicated, and it will not be evaluated explicitly here.
It is a form of the general structure:
\be
J_{\m 3} = c_1 A_{\m}^a d A^b A^c d^{abc} +
c_2 A_{\m}^a d K^b A^c d^{abc} + \cdots
\ee
invoving $A$, $K$,  $d$ and invariant tensors.
Here $c_i$ are numerical coefficients
and there are about a dozen of them.
There are at least four ways to construct
the polynomials $J$ and $C$ discussed here:
\begin{enumerate}
\item
Brute force--choose the most general possible solution with
arbitrary coefficients
and then fix the coefficients by algebra using
(\ref{omega}) and  (\ref{coveq}).
\item
Cartan Homotopy operators-find a solution to the homotopy equations
as outlined in \ci{msz} and \ci{zumino}, as was done
in \ci{dds}, and then integrate over t.
\item
Construct the covariant derivative using the Hamiltonian methods
given in \ci{bergs} with the Lagrangian given in \ci{dds}.
\item
Fock space homotopy--use the methods of\ci{dix}.
\end{enumerate}
All of these are quite labor intensive.
For many purposes it is probably sufficient
to know that the expressions
exist and satisfy the relevant equations--frequently a long
expression conveys less useful information than the knowledge
that it satisfies certain defining equations.
If actual construction is needed, the second method is probably
the best, and usually the answer should
probably be left in  integral form like (\ref{MM}), since that
form preserves much of the information.

\section{Commutators of the Covariant Derivatives}
We know that
\be {\cal D}_{ \m \n }  =
 [{\cal D}_{\m}
,  {\cal D}_{\n} ]
\ee
commutes with the BRS operator, and therefore should
represent some sort of curvature.  We find for the
string, setting $V^{\m}(\s)$ in
(\ref{string}) equal to  $\d^{\m}_{\n}$,  that:
\[
{\cal D}_{\m \n\; {\rm string}} =
\prod_{\s' \n, m} \int dX^{\n}(\s') dy^m(\s')
\]
\be
\left \{
\left (
\int d \s \left [
F_{\m \n}^a T_a(\s)
- 6  [ Q[A]_{\m \n \l} + \pa_{[\m} B_{\n \l]} ]
 \fr{ d X^{\l} }{ d \s} \right ] \F
\right )
 {\d \over \d \F} \right \}
\label{unexpected}
\ee
where $ Q[A]_{\m \n \l}$ is some unfamiliar function of
$A$.  This is not what one might naively expect, since it
does not involve the invariant
\be
H_3^0 = d B^0_2 + I^0_3
\label{invat}
\ee
 However we can add
a term to the covariant derivative:
\be
{\cal D}_{\m \; {\rm new}}
= {\cal D}_{\m \; {\rm old}}
-{1 \over 2} \int \; d^Dx \; \left \{
{F}^a_{\m \n} \fr{\d}{\d A^a_{\n}} \right \}
\label{newercov}
\ee as was done in equation (\ref{newcov}).
Here by $ {\cal D}_{\m \; {\rm old}}$ we mean
(\ref{gencov}). Using
(\ref{newercov}) we
do get a   result close to the naively expected one:
\be
{\cal D}_{\m \n\; {\rm new }} \F
= - 6 \int d \s
H^0_{\m \n \l}
 \fr{ d X^{\l} }{ d \s} \F
\label{expected}
\ee
Since the central extended $T^a$ term is no longer present here,
a zero commutator with $\d$ demands that the familiar invariant
(\ref{invat}) must appear.

The same trick will work also for the
p-branes, but there the invariants obtained by commutation
may also depend on $K$, since ${\cal D}_{\m}$ depends on $K$.

\section{Conclusion}

This paper has introduced two new polynomial
functions of $A$ and $K$ that appear in general when
one couples p-branes (with odd p) to Yang-Mills theory.
The polynomial
$C^1_p[A, \w, K]$  is needed to define the nilpotent
BRS operator $\d$ for the general p-brane.
It is simply   related (by descent equations) to the polynomial
$C^0_{p+1}[A, K]$ that appears in the
WZW part of the p-brane action in \ci{dds}.
The polynomial
$J^0_{\m p}[A, K]$ is needed in the covariant derivative
${\cal D}_{\m}(\s)$
to ensure that $[ \d, {\cal D}_{\m}(\s)] = 0$,
and a proof that it can always be constructed has been given
for all odd p.

In summary, for all odd $p \geq 1$, the BRS transformation
of the p-brane field $\F$
takes the form
\be \d   \F  =
   \int  \left [ -
   \w^a T^a(\s)  d^p \s
+ n C^1_p[ A, K](\s)
+ \L_p^1(\s)  \right ] \F
\ee
where the p-form $C^1_p(A, K,\w)$ satisfies
the relation:
\be
\d C^1_p(A, K,\w)   = I^2_p(K,\w)  -   I^2_p(A,\w)
+ d C^2_{p -1}(A, K,\w)
\ee
The general Mickelsson-Faddeev algebra has
been discussed.  It is determined by the ghost
charge two term in the descent equations:
\be
\int_{\s} I^2_p(K,\w)(\s)
= \int_{\s} \left (  \w^a d \w^b M_{p-1}^{(ab)}[K] \right )
= \int_{\s} \left (  d \w^a d \w^b N_{p-2}^{(ab)}[K] \right )
\ee
where the latter ($N$) form is valid for $p \geq 3$.
Then the covariant derivative takes the form:
\be
 {\cal D}_{\m}(\s) \F  =
  \left \{
\fr{\d}{\d X^{\m}(\s)} + A_{\m}^a(\s) T_a(\s)
 +      n  J^0_{\m p }(\s) - (p+1)  B_{\m p}(\s)
   \right \}   \F
\ee
where the function $J^0_{\m p}$ is determined by:
\be  [ \d ,  J^0_{\m p }(\s)]=  J^1_{\m p }(\s)
 \ee
with $J^1_{\m p }$ a calculable function determined by
$C^1_p(A, K,\w)$ and $I^2_p(K,\w)(\s)$.

In conclusion,
the bosonic part of the
loop space BRS operator and the loop space covariant derivative
of the string generalize to the p-branes in a fairly simple way, as
did the bosonic part of the action coupled to background Yang-Mills fields.
It remains to be seen whether fermions with heterosis and $\k$ symmetry can
also be incorporated.

{\bf Acknowledgment}  I have enjoyed useful conversations with
Mike Duff and Ergin Sezgin.

\section{ Appendix A: Construction of $J^0_p$.}

 {\bf Theorem}

{ \em The local unintegrated BRS cohomology of the nilpotent BRS
operator defined by:
\be
  \d A_{\m}^a
=
D(A)_{\m}^{ab} \w^b =
\pa_{\m} \w^a + f^{abc} A_{\m}^b \w^c
\label{afirst}
\ee
\be
\d
K_{i}^a(y)
=
D(K)_{i}^{ab} \w^b =
\pa_i X^{\m} \pa_{\m} \w^a + f^{abc} K_i^b \w^c
\ee
\be
\d \w^a(x)
=
- \fr{1}{2} f^{abc} \w^b \w^c
\label{alast}
\ee
consists of sums of products of functions of the form
\be
 I[\pa_{\r}, A_{\m}^a, K_i^b,\pa_j X^{\n}] T^{g}(\w),
\ee
where the terms are separately invariant:
\be
\d I[\pa_{\r}, A_{\m}^a, K_i^b,\pa_j X^{\n}]   =0
\ee
\be
\d T^{g}(\w)=0
\ee
}
Here the superscript $g$ is the ghost number of the corresponding expression.
In other words all solutions of
\be
\d P^g = 0
\ee
can be expressed in the form
\[
P^g = \d P^{g-1}
\]
\be
+ I[\pa_{\r}, A_{\m}^a, K_i^b,\pa_j X^{\n}]
T^{g}(\w),
\ee
where $P^{g}$ are unintegrated local polynomials
that depend only on the fields $A,K,\w$ and
the derivative operator.
Now for a semisimple group, the
polynomials $T^g(\w)$ exist
only for $g=3,5,7...$ and not for
$g=1$, which is all we need for present purposes.
In particular, since there are (for semisimple
groups)
no $T^g$ for $g=1$, it follows that every polynomial
$J^1$ of ghost charge one which satisfies:
\be
\d J^1[ A_{\m}^a, K^a_i ]
 = 0
\ee
can be written in the form:
\be
J^1 = \d J^{0}[ A_{\m}^a, K^a_i ]
\ee
where
$ J^{0}[ A_{\m}^a, K^a_i ] $ is another  local polynomial.
This is the result used in the text to show that a polynomial
$J^0_{\m p}$ satisfying (\ref{coveq}) exists for all odd p.

{\bf Sketch of Proof:}

This result is
easily proved along the lines given in
\ci{dix} using the following field redefinitions:
\be
\f^a_i = \Pi^{\m}_i A^a_{\m} - K^a_i
\ee
which give rise to  transformations like (\ref{brsf})
in the introduction.
Note that since here we do not have an integration over
spacetime, we do not need to consider the full operator
with an exterior derivative term
$\d + d$ here but only the part $\d$.
We then use a spectral sequence based on the counting operator:
\be
N = N(A) + N(\w) + N(\f)
\ee
This means that
\be
\d_0 = \int d^D x \pa_{\m} \w^a {\d \over \d A_{\m}^a}
\ee
So the space $E_1$ is of the form:
\be
E_1 = E_1[ A', \f, \w]
\ee
where it is understood that only undifferentiated
$\w$ appears in this expression.  Next
\be
d_2 = \P_1 \w^a \left [ T^a - {1\over2} Y^a  \right ] \P_1
\ee
Then the theorem is
a direct consequence of the arguments in \ci{dix}, and
the demonstration is in fact much easier here than in
that case because there
is no exterior derivative here to give rise to
exceptional vectors which require   the
consideration of $d_r$ for $r\geq 2$.
The spectral sequence collapses at $E_2 = {\rm ker}\; \D_1$
and the result is as stated.

\end{document}